\documentclass{article}
\usepackage{stywhispers,amsmath,epsfig}
\usepackage{setspace}
\usepackage{geometry} 
\usepackage{epstopdf}
\usepackage{tikz}
\usepackage[labelsep=period]{caption}
\newcommand\copyrighttext{%
  \footnotesize \copyright~2019 IEEE. Personal use of this material is permitted. Permission from IEEE must be obtained for all other uses, in any current or future media, including reprinting/republishing this material for advertising or promotional purposes, creating new collective works, for resale or redistribution to servers or lists, or reuse of any copyrighted component of this work in other works.}
\renewcommand\copyrightnotice{%
\begin{tikzpicture}[remember picture,overlay]
\node[anchor=south,yshift=10pt] at (current page.south) {\fbox{\parbox{0.7\dimexpr\textwidth-\fboxsep-\fboxrule\relax}{\copyrighttext}}};
\end{tikzpicture}%
}

\usepackage{siunitx}
\usepackage{microtype}
\usepackage{etoolbox}
\robustify\bfseries
\usepackage{upgreek}
\usepackage[hidelinks]{hyperref}
\usepackage{tabularx}
\usepackage{booktabs}
\usepackage{subfig}
\usepackage{multirow}
\usepackage{color}
\usepackage{paralist}
\usepackage{amsmath}
\usepackage[super]{nth}
\usepackage[nameinlink]{cleveref}
\usepackage{float}
\usepackage{stfloats}
\usepackage{soul}
\title{Application of different simulated spectral data and Machine learning to estimate the Chlorophyll a Concentration of several inland waters}

\name{Philipp M. Maier, Sina Keller\thanks{We thank the German Federal Ministry of Education and Research for funding the WAQUAVID project.}}
\address{Karlsruhe~Institute~of~Technology~(KIT)\\
	Institute~of~Photogrammetry~and~Remote~Sensing~(IPF),\\
	Englerstra{\ss}e 7, 76131 Karlsruhe, Germany}

\begin{document}

\maketitle

\copyrightnotice

\begin{abstract}
Water quality is of great importance for humans and for the environment and hence has to be monitored continuously.
One possibility are proxies such as the chlorophyll~\textit{a} concentration, which can be monitored by remote sensing techniques.
This study focuses on the trade-off between the spatial and the spectral resolution of six simulated satellite-based data sets when estimating the chlorophyll~\textit{a} concentration with supervised machine learning models.
The initial dataset for the spectral simulation of the satellite missions contains spectrometer data and measured chlorophyll~\textit{a} concentration of $13$ different inland waters.
The analysis of the regression performance indicates, that the machine learning models achieve almost as good results with the simulated Sentinel data as with the simulated hyperspectral data.
Regrading the applicability, the Sentinel~2 mission is the best choice for small inland waters due to its high spatial and temporal resolution in combination with a suitable spectral resolution.

\end{abstract}


\section{Introduction}
\label{sec:intro}

According to the sixth sustainable development goals set by the United Nations in 2018, clean water is a key resource for humans and the environment~\cite{SDG_UN_2018}.
However, the water quality is threatened extensively by human influences such as emission of wastewater or overfertilization caused by agriculture.
Hence there is a great demand for a continuous and efficient system to monitor water quality (cf.~\cite{Koponen.2002, PalmerS.KutserT.HunterP.D..2015, MaierP.M.HinzS.KellerS..2018}).

In addition to commonly applied in-situ probes, remote sensing as a technique is often considered when monitoring large water surfaces.
Remote sensing offers some advantages over point sample measurement.
In particular, satellite image data is frequently available and it is cost-efficient in the long run.
Furthermore, information about water quality parameters derived by satellite images are more representative than in-situ measured point values in terms of area-wide coverage. 

One important water quality parameter is chlorophyll~\textit{a} (chl~\textit{a}).
It serves as a proxy for the nutrition supply of a water body.
Chl~\textit{a} is a pigment which appears in phytoplankton, and provides the basis for photosynthesis.
The occurrence of phytoplankton depends on the natural nutrition supply of a water body as well as human sources. 

Chl~\textit{a} is detectable by passive remote sensing sensors in the visible spectrum.
An absorption feature in the spectral band region around \SI{665}{\nm} indicates chl~\textit{a}~\cite{MorelA.PrieurL..1977}.
Several studies have already demonstrated the applicability of remote sensing data with respect to the estimation of chl~\textit{a} concentrations in inland waters \cite{Koponen.2002, KellerS.MaierP.RieseF.NorraS.HolbachA.BorsigN.WilhelmsA.MoldaenkeC.Zaak.2018}.
To estimate the chl~\textit{a} concentration with spectral data, two complementary approaches are applied. 
First, engineering approaches consider spectral features or band ratios~\cite{GITELSON.1992, Gons.1999}.
Second, machine learning (ML) approaches have been emerged in the last decade~\cite{KeinerL.E.YanX.H..1998, Matarrese.2008, GonzalezVilas.2011, MaierP.M.KellerS..2018, KellerS.MaierP.RieseF.NorraS.HolbachA.BorsigN.WilhelmsA.MoldaenkeC.Zaak.2018}.
These approaches estimate the chl~\textit{a} concentration primarily in a supervised way without prior-knowledge of the underlying physical processes. 

In general, the estimation of chl~\textit{a} concentrations in water bodies from remote sensing data is a challenging task.
Inland waters are optically complex since they contain suspended and particular materials.
These materials are characteristic for every inland water~\cite{PalmerS.KutserT.HunterP.D..2015}.

Another limiting factor when monitoring inland waters is the spatial resolution of the satellite images.
Unfortunately, high spectral resolution is often accompanied by a lower spatial resolution.
In case of the oceans, this is not an issue.
With respect to inland waters however, the spatial resolution is crucial and hence an exclusion criteria of some satellite sensors.
For example, the SeaWiFS (Sea-viewing Wide Field-of-view Sensor) as an ocean water observation satellite mission has a spatial resolution of more than \SI{1}{\kilo\meter}~\cite{OReillyJ.E.MaritorenaS.MitchellB.G.SiegelD.A.CarderK.L.GarverS.A.Kahru.1998}. 
Therefore, most of the smaller inland water bodies are represented by only one, mixed pixel which hinders the use of satellite data for the estimation of the chl~\textit{a} concentration of small water bodies. 

Some studies investigate the trade-off between spectral and spatial resolution of satellite data recorded by the common missions~\cite{Decker.1992, Beck.2016}.
A thorough analysis of the estimation performance of feature engineering approaches on chl~\textit{a} concentrations for several simulated satellite sensors is presented in~\cite{Beck.2016}.
Previous work~\cite{MaierP.M.KellerS.2019} addresses the effect of different hyperspectral resolutions of the input data and machine learning models when estimating chl~\textit{a} concentrations.

In this study, we simulate satellite data with respect to several multi- and hyperspectral satellite missions such as Landsat~5, Landsat~8, Sentinel~2, Sentinel~3, EnMAP and Hyperion. 
The basis of the simulated data is a spectrometer dataset of $13$ different inland waters which was conducted in the surrounding region of Karlsruhe (Germany) during the summer 2018.
In total, the dataset contains $408$ datapoints.
Each datapoint consists of the spectral information and the associated chl~\textit{a} concentration.
The simulated spectral data serves as input data for selected ML models to estimate the chl~\textit{a} concentration of the different inland waters. 

The objectives of this contribution are:
\begin{compactitem}
    \item the simulation of satellite data based on the measured spectrometer data by applying the spectral response function or a Gaussian function (\Cref{sec:format});
    \item the estimation of the chl~\textit{a} concentration by applying different supervised ML models such as random forest (RF), support vector machine (SVM), multivariate adaptive regression spline (MARS) and an artificial neural network (ANN) on the respective simulated data (\Cref{sec:typestyle});
    \item the comparison of the regression performance in terms of simulated data and applied ML model (\Cref{sec:typestyle});
    \item the discussion of the regression performance with the focus on the spectral and the spatial resolution of the input data (\Cref{sec:typestyle}).
\end{compactitem}

\section{Dataset and data simulation}
\label{sec:format}

The data used in this contribution is from a measurement campaign~\cite{MaierP.M.KellerS.2019} in the surroundings of Karlsruhe, a city located in the Southwest of Germany.
During the summer of 2018, $13$ different inland water bodies were measured with a spectrometer and water samples were evaluated with a photometer.
A detailed description of the measurement campaign including the measurement setup is given in~\cite{MaierP.M.KellerS.2019}.

The spectrometer records hyperspectral data in a spectral range of \SI{341}{\nm} to \SI{1015}{\nm} with a sampling interval of~\SI{0.66}{\nm}.
Its measurement principle is based on the ratio between the incoming and the up-welling radiance in the perpendicular direction.
The spectrometer was mounted on a tripod, which was placed as far as possible in the water in case of a natural water body. When measuring an artificial water body, the spectrometer was set outside the water.

The water samples for the chl~\textit{a} concentration analysis, which we use as reference data, were collected close to the spectrometer.
The measured chl~\textit{a} concentrations and the respective spectra of the continuous spectrometer measurements were matched by their respective timestamps.
In total, we obtain a dataset with $408$ datapoints. 
Each datapoint consists of the spectral data and a chl~\textit{a} concentration value.  

\begin{figure*}[tb]
	\centering
    \includegraphics[width=0.9\textwidth]{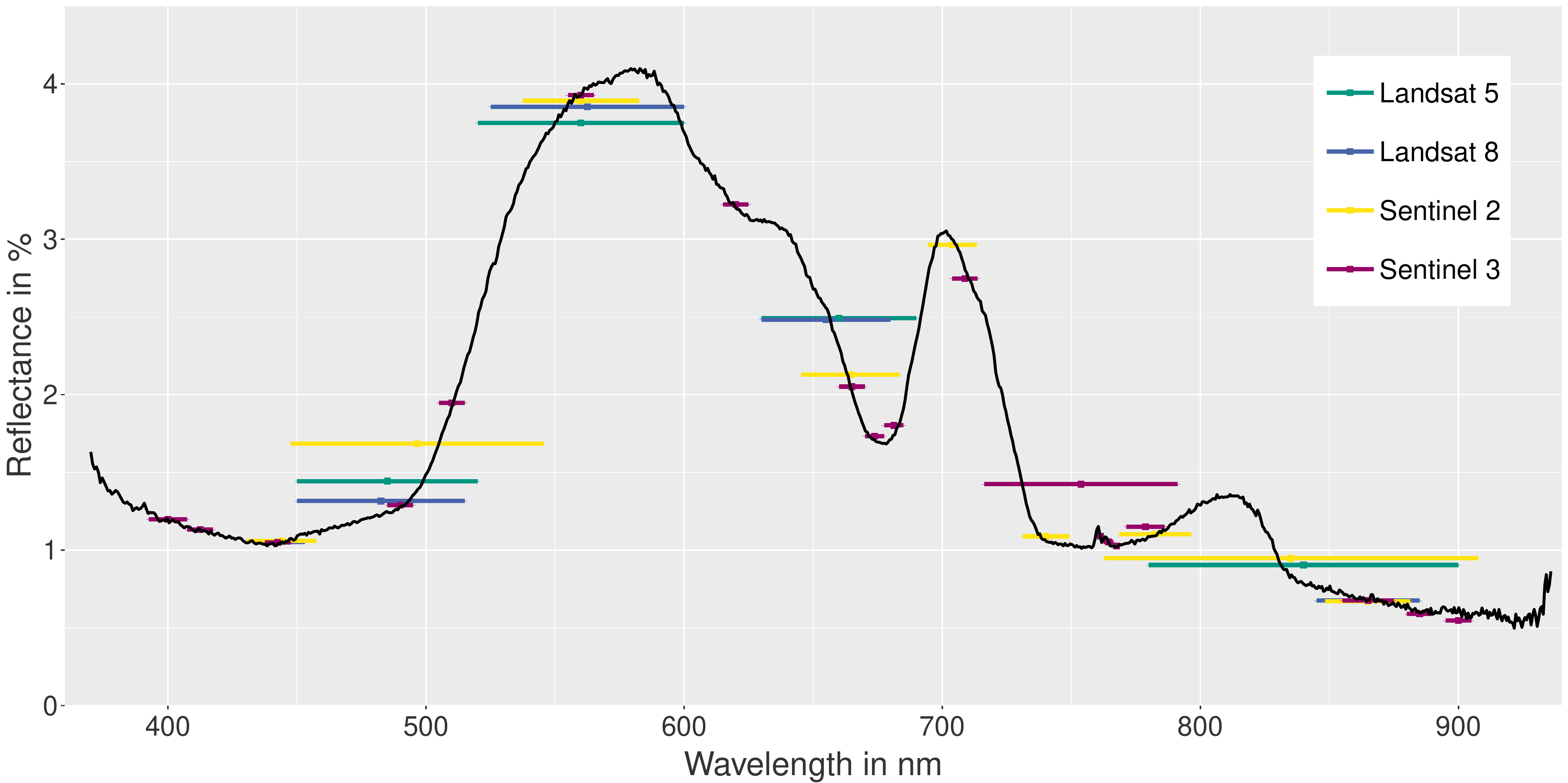}
    \caption{Median spectra of the spectrometer dataset and symbolization of the width of the satellite bands (colored lines). The dots in the middle of each bandwidth represent the simulated reflectance value of the band.}
    \label{fig:bands}
\end{figure*}

For our satellite-based simulation of the spectral data, we used spectrometer data in the wavelengths range of \SIrange{400}{900}{\nm} is used. 
The simulation of the spectra in accordance with the satellite missions was conducted with the hsdar-package in R~\cite{LehnertL.W.MeyerH.BendixJ}.
Three different approaches exist to calculate the satellite bands out of spectral data with different weighting functions: a Gaussian function, an equal-weighted function and the actual spectral response function. 
To calculate the spectra according to the Sentinel~2, Landsat~5 and Landsat-8 missions, we relied on the real spectral response function.
When simulating Sentinel~3, the EnMAP and Hyperion satellite missions, we applied the Gaussian function.
In the case of Sentinel~3, which is not implemented in the hsdar-package, we used the parameters central wavelength and full width at half maximum according to~\cite{Fletcher.} and a Gaussian function to simulate the bands.
\Cref{tab:Overview} gives an overview of the spectral and spatial characteristics of the satellite missions which have been used for the data simulation. 
Furthermore, \Cref{fig:bands} illustrates the bandwidth of each satellite mission in the spectral range of \SIrange{400}{900}{\nano\meter}.

\begin{table*}[tb]
    \centering    
    \caption{Summary of some characteristics of the different satellite systems used for the data simulation covering the spectral range between \SIrange{400}{900}{\nm}. The hyperspectral satellite missions are highlighted by ${\ast}$.}
    \begin{tabular}{lcccccc}
        \toprule
        {Satellite}& {Number} & {Bandwidth} & {Spectral range} & {Spatial resolution} & {Approach for} & {Data}\\
        {mission} & {of bands} & {in nm} & {in nm} & {in m} & {the simulation} & {source}\\
        \midrule
        Sentinel~2 & $9$ & {\numrange{18}{145}} & {\numrange{443}{865}} & {\numrange{10}{60}} & {Response function} & \cite{LehnertL.W.MeyerH.BendixJ} \\
        Sentinel~3 & $19$ & {\numrange{2.5}{75}} & {\numrange{400}{900}} & {\numrange{300}{1000}} & {Gaussian function} & \cite{Fletcher.} \\ 
        Landsat~8 & $5$ & {\numrange{16}{60}} & {\numrange{443}{865}} & $30$ & {Response function} & \cite{LehnertL.W.MeyerH.BendixJ}\\
        Landsat~5 & $4$ & {\numrange{60}{140}} & {\numrange{485}{840}} & $30$ & {Response function} & \cite{LehnertL.W.MeyerH.BendixJ}\\
        Hyperion$^{\ast}$ & $54$ & $10$ & {\numrange{406}{895}} & $30$ & {Gaussian function} & \cite{LehnertL.W.MeyerH.BendixJ}\\
        EnMAP$^{\ast}$ & $77$ & $6.5$ & {\numrange{423}{895}} & $30$ & {Gaussian function} & \cite{LehnertL.W.MeyerH.BendixJ}\\
        \bottomrule
    \end{tabular}
    \label{tab:Overview}
\end{table*}

\section{Methodology}
\label{sec:pagestyle}

For the estimation of the chl~\textit{a} concentration based on the different simulated satellite data, we selected four ML models: support vector machine~\cite{Vapnik.2013}, random forest~\cite{Breiman.2001}, multivariate adaptive regression spline~\cite{Milborrow.2018} and an artificial neural network~\cite{Ripley.1996}.
The applied ML models are inspired by the selection in~\cite{MaierP.M.KellerS..2018} due to their satisfactorily performance. 

To apply these models, the dataset consisting of the chl~\textit{a} values and the simulated satellite data was prepared. 
It was split into five equally sized parts with respect to distribution of the target variable, the chl~\textit{a} concentration.
Then, each of those parts was split randomly into two subsets: a training subset and a test subset. 
All five training subsets were aggregated to the final training subset. 
The test subset was generated similarly.
As a result, the distribution of the chl~\textit{a} concentration in the training as well as the test subset were representative compared to the reference measurements.

The training subset was used for the training of the ML models, while the test subset remained unused until the test phase. Before starting with the training, we applied a grid search to adjust the hyperparameters of the models. 
For example, hyperparameters of the SVM model are the penalty function cost and the kernel parameter gamma.

During the test phase, the models were validated on the yet unknown test dataset. 
The performance of the regression was expressed by the coefficient of determination ($R^2$) and the mean absolute error (MAE).  
Following the regression performance on the same database in~\cite{MaierP.M.KellerS.2019}, we also calculated the first derivative of the spectra for the simulated hyperspectral data of the Hyperion and EnMAP mission and applied those derivatives as input data for the RF and MARS model. 
In addition, we pre-processed the simulated satellite data with a scaling to ensure good regression results for the the MARS, SVM and ANN models.

\section{Results and Discussion}
\label{sec:typestyle}

\begin{figure*}[tb]
	\centering
    \includegraphics[width=0.85\textwidth]{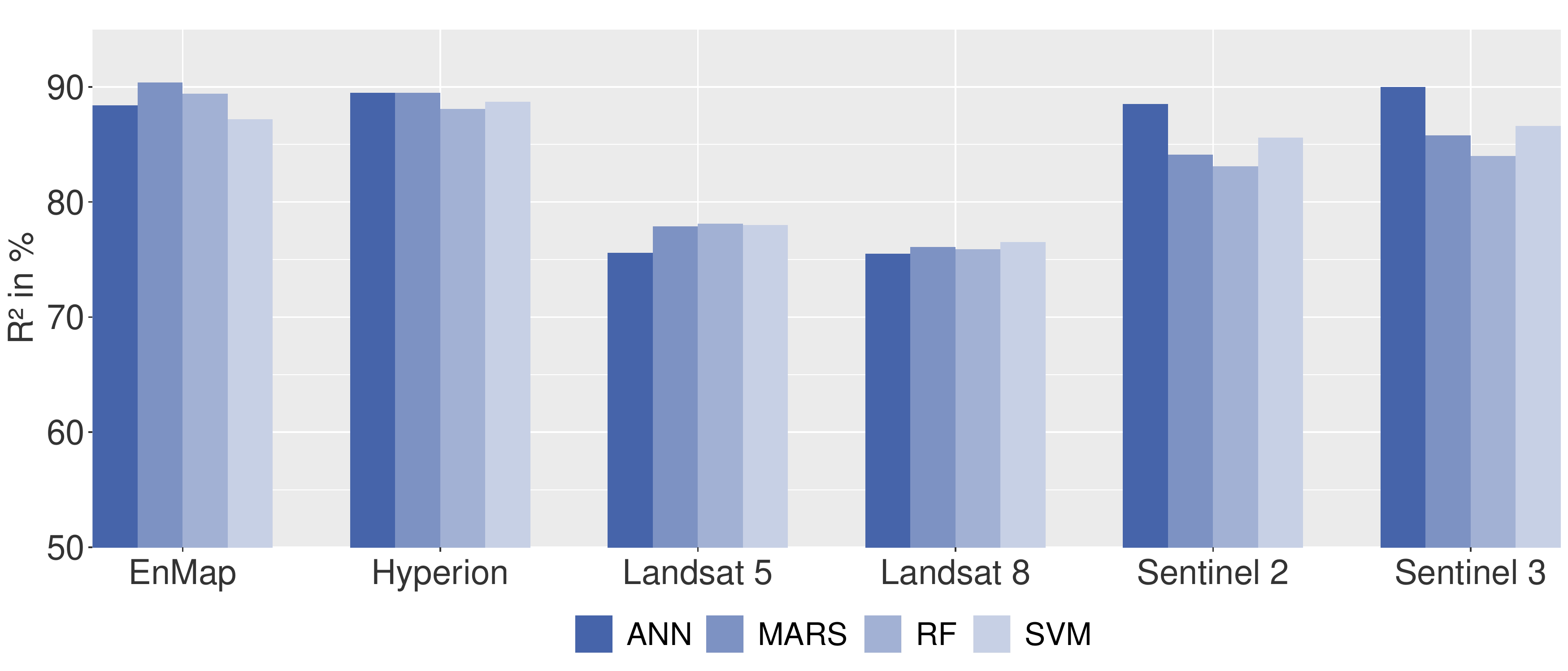}
    \caption{Regression results ($R^2$ in~$\%$) of the four ML models with different simulated satellite data.}
    \label{fig:bars}
\end{figure*}

\Cref{fig:bars} and \Cref{tab:MAE} present the regression performance of estimating the chl~\textit{a} concentration with respect to the applied ML models as well as the different simulated satellite data.
Regarding \Cref{fig:bars}, the regression performance of the four ML models are in the same range. 

When considering the simulated satellite input data for estimating the chl~\textit{a} concentration, the regression results expressed as $R^2$ are distinguishable. 
For the simulated hyperspectral satellite data (EnMAP and Hyperion), the coefficient of determination ($R^2$) is quite similar.
In case of the simulated Landsat data, the regression results are closely related. 
In detail, the ANN model performs worse than the other three models on these two simulated datasets. 
However, for the simulated Sentinel data, the ANN model provides the best regression results.

Considering the different simulated satellite data, the regression with the simulated hyperspectral data based on the EnMAP and Hyperion mission achieves the best results. 
The corresponding MAE values range between \SIrange{10.1}{12.6}{\micro\gram\per\liter}.
The MAE values of the models with simulated multispectral data according to the Sentinel missions is in the range between \SIrange{10.9}{14.8}{\micro\gram\per\liter}.
The estimation of the chl~\textit{a} concentration of all regression models with simulated Landsat data performs the worst compared to the other simulated satellite data. 
The MAE ranges between \SIrange{17.8}{20.5}{\micro\gram\per\liter}.
 
Analyzing bandwidth, number of bands, spectral range and resolution of the simulated satellite data, \Cref{fig:bands} shows that Landsat~5 (green) and Landsat~8 (blue) have similar bands with a similar band positioning.
The three bands between \SIrange{450}{700}{\nm} are nearly the same.
In the spectral range of \SIrange{800}{900}{\nm} Landsat~8 provides a narrower band than Landsat~5 and it has an additional fifth narrow band near \SI{430}{\nm}.
With respect to the estimation of the chlorophyll \textit{a} concentration, this additional band has no further impact on the regression task.

Similar to the simulated Landsat data, the simulated multispectral Sentinel~3 data provides a better spectral resolution and accounts for more bands with narrower bandwidths than Sentinel~2.
However, the regression performance of the ML models on simulated Sentinel~3 data is not clearly better than the regression performance of the models with simulated Sentinel~2 data.
When comparing the estimation performance with either simulated Sentinel data or simulated Landsat data, the outperformance of the models using the simulated Sentinel data can be well explained.
First, the simulated Sentinel data is characterized by more bands.
And second, these bands are well positioned within the spectral range of \SIrange{400}{900}{\nano\meter}.
For example, the simulated Sentinel data includes the extremes in the range of \SIrange{660}{710}{\nm} which are related to chl~\textit{a}.
The mentioned spectral range is not included in the two Landsat missions and explains the poor chl~\textit{a} estimation of all models~\cite{Decker.1992}.

The simulated hyperspectral data (EnMAP and Hyperion) with a nearly constant spectral resolution of \SI{6.5}{\nm} and \SI{10}{\nm} are not shown in \Cref{fig:bands} due to reason of transparency.
Comparing the regression results with the simulated hyperspectral and the simulated Sentinel data, the models relying on the hyperspectral datasets perform only slightly better.
This finding indicates that the band positioning of the Sentinel missions is good for the estimation of chl~\textit{a} concentrations.

Regarding the applicability of the simulated satellite data for a general monitoring approach in the context of inland waters, the Sentinel~2 data serves its purpose.
It provides data with appealing spectral resolution, a sufficient spatial resolution and is characterized by a high temporal frequency.
Hyperspectral data with a better spectral resolution leads to a satisfying  chl~\textit{a} estimation by applying the same ML models. 
However, their temporal resolution stays behind the temporal resolution of the Sentinel missions referring to two satellite systems.
Differentiating between the two Sentinel missions, the application of the Sentinel~3 satellites is limited to large inland water surface due to their poor spatial resolution of \SIrange{300}{1000}{\meter}.
In addition, the Landsat satellite missions provide an attractive spatial and temporal resolution as well.
However, the regression results of the models are the worst with this data since the Landsat missions are characterized by the lowest spectral resolution of all simulated satellite missions.

\renewcommand{\arraystretch}{1.0}
\begin{table}[tb]
	\centering
	\caption{Performance of the regression models expressed by MAE in \si{\micro\gram\per\liter}.}
	\resizebox{\linewidth}{!}{
		\begin{tabular}{lSSSS}
		\toprule
			\multirow{1}{*}{Simulated satellite data} & {RF} & {SVM} & {ANN} & {MARS}\\
            \midrule
			EnMAP		& {10.9}  &   {12.6} & {11.7} & {10.1}\\
            Hyperion 	& {11.3}  &   {12.2} & {11.3} & {10.5}\\
            Landsat~5		& {17.8}  &   {18.5} & {19.6} & {19.0}\\
            Landsat~8     & {18.8}  &   {18.8} & {20.0} & {20.5}\\
            Sentinel~2     & {14.8}  &   {13.2} & {11.5} & {14.2}\\
            Sentinel~3     & {14.3}  &   {14.1} & {10.9} & {13.0}\\
             \bottomrule
		\end{tabular}}
	\label{tab:MAE}
\end{table}

\section{Conclusion}
\label{sec:majhead}

In this paper, we address the estimation of chl~\textit{a} concentration with different simulated spectral data and supervised ML models. 
We rely on a spectrometer dataset measured at several inland water bodies.
For the simulation of the satellite-base data, we chose six different satellite missions as examples.
In addition, we apply four different supervised ML models for the estimation of the chl~\textit{a} concentration.

When comparing the simulated satellite data, the regression performance of all models with the simulated hyperspectral data achieves the best results due to their spectral and spatial resolution.
Referring to the estimation results, the ML models combined with the simulated Sentinel data are slightly worse than the estimation based on the simulated hyperspectral data.
Regarding the applicability for a generic monitoring approach of inland waters, the Sentinel~2 mission provides the best option for smaller water bodies.
The Sentinel~3 mission poses an alternative for large water bodies.

When focusing on the different ML models, the choice of a specific ML model has a minor impact on the regression performance.
Solely, the ANN models outperforms the other models when using the simulated Sentinel data.

In this study, we have focused on the estimation of the the chl~\textit{a} concentration as a selected water quality parameter. 
For the estimation of further quality parameters such as different algae types, the (simulated) hyperspectral data could provide an excellent basis due 
to its high spectral resolution. 
The choice of ML models and the (simulated) satellite data has to be adapted according to the respective water quality parameter which will be estimated.  
This investigation could be addressed in future work.

\bibliographystyle{IEEEbib}

\bibliography{Literatur}

\end{document}